\documentclass[preprint, showpacs,preprintnumbers,amsmath,amssymb,nofootinbib]{revtex4}
\usepackage{amssymb}
\usepackage{amsfonts}

 \usepackage{epsf}
 \usepackage{graphicx}
\usepackage{float}
\begin{document}
\newcommand{\be}[1]{\begin{equation}\label{#1}}
 \newcommand{\ee}{\end{equation}}
 \newcommand{\bea}{\begin{eqnarray}}
 \newcommand{\eea}{\end{eqnarray}}
 \newcommand{\bed}{\begin{displaymath}}
 \newcommand{\eed}{\end{displaymath}}
 \def\disp{\displaystyle}

 \def\gsim{ \lower .75ex \hbox{$\sim$} \llap{\raise .27ex \hbox{$>$}} }
 \def\lsim{ \lower .75ex \hbox{$\sim$} \llap{\raise .27ex \hbox{$<$}} }

\title{Unbiased constraints on the clumpiness of the Universe from standard candles}

\author{Zhengxiang Li$^{}\footnote{Electronic address: zxli918@bnu.edu.cn}$,
        Xuheng Ding$^{}\footnote{Electronic address: dingxuheng@mail.bnu.edu.cn}$, and
        Zong-Hong Zhu$^{}\footnote{Electronic address: zhzhu@bnu.edu.cn}$}

\address{Department of Astronomy, Beijing Normal University,
          Beijing 100875, China}

\begin{abstract}
We perform unbiased tests for the clumpiness of the Universe by
confronting the Zel'dovich-Kantowski-Dyer-Roeder luminosity
distance,  which describes the effect of local inhomogeneities on
the propagation of light with the observational one estimated from
measurements of standard candles, i.e., type Ia supernovae (SNe Ia)
and gamma-ray bursts (GRBs). Methodologically, we first determine
the light-curve fitting parameters which account for distance
estimation in SNe Ia observations and the luminosity/energy
relations which are responsible for distance estimation of GRBs in
the global fit to reconstruct the Hubble diagrams in the context of
a clumpy Universe. Subsequently, these Hubble diagrams allow us to
achieve unbiased constraints on the matter density parameter
$\Omega_m$, as well as the clumpiness parameter $\eta$ which
quantifies the fraction of homogeneously distributed matter within a
given light cone. At a 1$\sigma$ confidence level, the constraints
are $\Omega_m=0.34\pm0.02$ and $\eta=1.00^{+0.00}_{-0.02}$ from the
joint analysis. The results suggest that the Universe full of
Friedman-Lema\^{i}tre-Robertson-Walker fluid is favored by
observations of standard candles with very high statistical
significance. On the other hand, they may also indicate that the
Zel'dovich-Kantowski-Dyer-Roeder approximation is a sufficiently
accurate form to describe the effects of local homogeneity on the
expanding Universe.
\end{abstract}


 \maketitle
 \renewcommand{\baselinestretch}{1.5}

\section{INTRODUCTION}
The standard physical model of cosmology is based on the solution of
general relativity describing a spatially homogeneous and isotropic
spacetime, known as the Friedmann-Lema\^{i}tre-Robertson-Walker
(FLRW) solution. It is assumed that the geometry of our Universe is
smooth on large scales. One of the major tasks in modern cosmology
is to precisely determine the parameters which characterize the
postulated model by fitting the observational data. The cornerstone
of observational evidence that supports the FLRW model is the
existence of highly isotropic cosmic microwave background radiation
(CMBR). It could be inferred that the spacetime should be exactly
FLRW when the background radiation appears to be exactly isotropic
to a given family of observers~\cite{Ehlers68}. Therefore,we can
prove the Universe to be FLRW just from our own observations of the
CMBR by taking the Copernican principle into consideration.
Moreover, this result could be extended to the case of an almost
isotropic background radiation, which hints at an almost FLRW
spacetime~\cite{Stoeger95}. Although this simple solution of
Einstein field equation provides an excellent description for the
universe on large scales, it also makes clear that we need to
understand the departures from a spatially homogeneous model when
interpreting observational data. Indeed, departures from perfect
homogeneity change the distance-redshift relation. However, in
practice, cosmological observations are usually fitted just using
relationships derived from homogeneous models.

The fact that matter is not continuously distributed can imprint
most cosmological observations probing quantities related to light
propagation(as discussed in detail in Ref.~\cite{Clarkson12}), in
particular regarding the propagation of light with narrow beams,
such as the redshift, the angular diameter distance, the luminosity
distance, and the image distortion. The importance of quantifying
the effects of inhomogeneities on light propagation was first
pointed out by Zel'dovich~\cite{Zeldovich64} and
Kantowski~\cite{Kantowski69}. They designed an ``empty beam"
approximation by arguing that photons should mostly propagate in
vacuum. Later, this was generalized by Dyer and Roeder as the
``partially filled beam" approach~\cite{Dyer72,Dyer73}. More
generally, the early work of Ref.~\cite{Zeldovich64} stimulated many
studies on this issue~\cite{Dashevskii66,
Bertotti66,Gunn67a,Gunn67b,Refsdal70,Weinberg76,Dyer81,Linder88,Fang89,Wux90,Tomita98,Rose01,Kibble05}.
In this framework, the proportion of clumped matter with respect to
the homogeneous fluid is characterized by the clumpiness or
smoothness parameter. In addition, they arrived at an equation for
the angular diameter distance which, via the Etherington relation,
connects to the observable luminosity distance. We refer to it here
as the Zel'dovich-Kantowski-Dyer-Roeder (ZKDR) luminosity distance.

Since the 1960s, a rich literature has formed which concerns the
ZKDR approach and its cosmological implications. Phenomenologies and
investigations involving many different physical aspects were
performed, such as analytical or approximate
expressions~\cite{Kantowski98,Kantowski00,Demianski03}, critical
redshift for the angular diameter distance~\cite{Sereno01},
gravitational lensing~\cite{Covone05,Giovi01}, and accelerated
expanding Universe models driven by particle
creation~\cite{Campos04}. Recently, some quantitative analysis from
such compact radio sources as standard
rulers~\cite{Alcaniz04,Santos08a}, and such type Ia supernovae (SNe
Ia) or gamma-ray bursts (GRBs) as standard
candles~\cite{Santos08b,Busti11,Busti12,Breto13,Yang13,Lima14} were
also performed. To be specific, in Ref.~\cite{Santos08b},
constraints on the dark energy and smoothness parameter from the
so-called gold SN Ia sample released by the High-z Supernova
team~\cite{Riess07} and the first year results of the Supernova
Legacy Survey (SNLS), which is a planned five-year
project~\cite{Astier05}, were examined. The results suggested that
SNe Ia data alone was incapable of constraining the smoothness
parameter although the gold SN Ia provided a little more stringent
constraint since this sample extended to appreciably higher
redshifts. Later, Busti {\it et al.}~\cite{Busti12} performed an
updated investigation where the statistical analysis was based on
the 557 SNe Ia Union2 compilation data~\cite{Amanullah10} and 59
Hymnium GRBs~\cite{Hao10}, and almost the same conclusion was
achieved. More recently, this issue was also studied by using
Union2.1 SN Ia~\cite{Suzuki12} plus nine long GRBs in $1.55\leq
z\leq3.57$~\cite{Tsutsui12} and the constrained value of the
smoothness parameter indicated a clumped Universe~\cite{Breto13}. On
the other hand, as concluded in their work, this result may be an
indication that the ZKDR approximation is not a precise form of
describing the effects of clumpiness in the expanding Universe.

However, in these previous analysis, all distances of SNe Ia and
GRBs applied to test the inhomogeneity of the Universe were derived
from a global fit in the context of standard dark energy scenarios
where the clumpiness has vanished, i.e., the flat $\Lambda$ cold
dark matter ($\Lambda$CDM) or $w$CDM model. That is, the light-curve
fitting parameters accounting for the distance estimation in SNe Ia
observations (e.g., $\alpha$ and $\beta$ in the most widely used
SALT2 training method~\cite{Guy07}) are left as free parameters (on
the same weight as cosmological parameters) and are determined by
fitting the distances of SNe Ia, which is a linear combination of
light-curve fitting parameters and observed quantities, to the
model-predicted ones in the context of the standard $\Lambda$CDM or
$w$CDM scenario. Therefore, HDs constructed in this way are somewhat
model dependent. Moreover, cosmological implications on nonstandard
dark energy scenarios or a Universe with homogeneity taken into
consideration derived from these HDs are model
biased~\cite{Conley11}. It has been shown that this kind of bias
cannot be neglected and may be significant in the era of precision
cosmology~\cite{Zheng12,Zheng14}. Certainly, this kind of bias also
hides in the GRB cosmology where luminosity relations being
responsible for distance estimation of GRB are calibrated with the
model-dependent HDs of low-redshift SNe Ia~\cite{Liang08,Liang10}.

In this paper, we first reconstruct Hubble diagrams for the latest
SNe Ia and for long GRB observations by calibrating the light-curve
fitting parameters and luminosity relations, respectively, in the
context of an inhomogeneous Universe with the cosmological constant.
These Hubble diagrams can lead to unbiased tests for the matter
density parameter $\Omega_m$ as well as the clumpiness parameter
$\eta$. For the joint light-curve analysis of the SDSS-II and the
SNLS (JLA SN Ia) in the range of $0.01\leq
z\leq1.23$~\cite{Betoule14}, the constraints are
$\Omega_m=0.29^{+0.07}_{-0.05}$ and $\eta=0.76^{+0.24}_{-0.65}$,
slightly indicating a clumped Universe. For the long GRBs in the
range of $1.48\leq z\leq8.20$~\cite{Fayin11}, the best fits are
$\Omega_m=0.42\pm0.06$ and $\eta=1.00^{+0.00}_{-0.12}$, strongly
supporting a homogeneous Universe. For the combination of these two
probes, the constraints are $\Omega_m=0.34\pm0.02$ and
$\eta=1.00^{+0.00}_{-0.02}$, also favoring a universe full of FLRW
fluid with a very high confidence level. We suggest that the matter
density parameter $\Omega_m$ is mainly determined by the SNe Ia
observations while the clumpiness parameter $\eta$ is primarily
constrained from the observed GRB events. Moreover, it is also shown
that larger scales are explored, the test more strongly implies a
homogeneous Universe. These reasonable results may be an indication
that the ZKDR approximation remains to be a precise description for
the luminosity distance-redshift relation in a locally inhomogeneous
Universe with the cosmological constant.

\section{\bf THE ZKDR LUMINOSITY DISTANCE}
For most cosmological models, angular or apparent size distance,
which is proportional to the square root of the cross-sectional area
$A(z)$, is related to the luminosity distance by
$d_\mathrm{A}(z)=d_\mathrm{L}(z)/(1+z)^2$. In the model only
including dark matter and dark energy, the luminosity distance
$d_\mathrm{L}(z)$, which accounts for a partially depleted mass
density in the observing beam but neglects lensing by external
masses, is obtained by integrating the second-order differential
equation for $A(z)$ of an observing beam from the source at redshift
$z$ to the observer at $z=0$~\cite{Kantowski98,Kantowski01}:
\begin{equation}\label{eq1}
(1+z)^2E(z)\frac{d}{dz}\bigg[(1+z)^2E(z)\frac{d}{dz}\sqrt{A(z)}\bigg]+\frac{3}{2}\eta\Omega_m(1+z)^5\sqrt{A(z)}=0,
\end{equation}
where $E(z)$ is the reduced Hubble parameter at redshift $z$
\begin{equation}\label{eq2}
E(z)=\frac{H(z)}{H_0}=(1+z)\sqrt{1+\Omega_mz+\Omega_\Lambda[(1+z)^{-2}-1]},
\end{equation}
and the phenomenological parameter $\eta=1-\rho_{\mathrm{cl}}/\rho$
is the so-called clumpiness or smoothness parameter which quantifies
the amount of matter in clumps relative to the amount of matter
uniformly distributed. The required boundary conditions for
Eq.~(\ref{eq1}) are
\begin{equation}\label{eq3}
\sqrt{A}\mid_{z=0}=0,~~~~~~~~~~~~\frac{d\sqrt{A}}{dz}\mid_{z=0}=-\sqrt{\delta\Omega}\frac{c}{H_0},
\end{equation}
where $\delta\Omega$ is the solid angle of the beam. By using an
approximate change of variables
\begin{eqnarray}\label{eq4}
h(A,z)&\equiv&(1+z)\sqrt{\frac{A}{\delta\Omega}},\\
\zeta(z)&=&\frac{\Omega_m}{1-\Omega_m}(1+z)^3+1,
\end{eqnarray}
Eq.~(\ref{eq1}) can be transformed into a hypergeometric equation
\begin{equation}\label{eq5}
(1-\zeta)\zeta\frac{d^2h}{d\zeta^2}+\big(\frac{1}{2}-\frac{7}{6}\zeta\big)\frac{dh}{d\zeta}+\frac{\nu(\nu+1)}{36}=0.
\end{equation}
The resulting luminosity distance is then given by
\begin{equation}\label{eq7}
d_L(z)=(1+z)h(\zeta(0)).
\end{equation}
Expressed in terms of hypergeometric functions, Eq.~(\ref{eq7})
becomes
\begin{eqnarray}\label{eq8}
d_L(z;\Omega_m,\nu)=&\frac{c}{H_0}&\frac{2(1+z)}{\Omega_m^{1/3}(1+2\nu)}[1+\Omega_mz(3+3z+z^2)]^{\nu/6}\nonumber\\
&\times&\bigg\{~_2F_1\bigg(-\frac{\nu}{6},\frac{3-\nu}{6};\frac{5-2\nu}{6};\frac{1-\Omega_m}{1+\Omega_mz(3+3z+z^2)}\bigg)\nonumber\\
&\times&_2F_1\bigg(\frac{1+\nu}{6},\frac{4+\nu}{6};\frac{7+2\nu}{6};1-\Omega_m\bigg)\nonumber\\
&-&[1+\Omega_mz(3+3z+z^2)]^{-(1+2\nu)/6}~_2F_1\bigg(-\frac{\nu}{6},\frac{3-\nu}{6};\frac{5-2\nu}{6};1-\Omega_m\bigg)\nonumber\\
&\times&_2F_1\bigg(\frac{1+\nu}{6},\frac{4+\nu}{6};\frac{7+2\nu}{6};\frac{1-\Omega_m}{1+\Omega_mz(3+3z+z^2)}\bigg)\bigg\}.
\end{eqnarray}
The parameter $\nu$ presented in Eqs.~(\ref{eq5}) and~\ref{eq8}
corresponds to the clumpiness parameter $\eta$ by
\begin{equation}\label{eq9}
\eta=\frac{1}{6}(3+\nu)(2-\nu).
\end{equation}
The range for $\nu$ is $0\leq\nu\leq2$, where $\nu=0(\eta=1)$ is
related to a FLRW fluid, while $\nu=2(\eta=0)$ to a totally clumped
case.

Actually, the ZKDR approach has been criticized by several authors
(e.g., a few detailed comments gathered in Ref.~\cite{Breto13}).
However, so far, confrontations of the ZKDR luminosity distance with
observations have not led to conclusive results in the sense of
totally excluding this model. Moreover, we should keep in mind that
most previous tests in this field were somewhat dependent on the
standard dark energy model (the flat $\Lambda$CDM or $w$CDM).
Therefore, it is necessary to clarify the validity and the scope of
the ZKDR luminosity distance in describing the Universe in a
model-unbiased way. Here, we follow the simplest treatment, where
$\eta$ is assumed to be a constant.

\section{\bf SAMPLES AND RESULTS}
We carry out analysis by using the latest observations of standard
candles, including the joint light-curve analysis of the SDSS-II and
SNLS supernova samples~\cite{Betoule14}--which is referred to as JLA
SN Ia in the literature--and the long gamma-ray bursts reported in
Ref.~\cite{Fayin11}. Descriptions for the samples, methodology, and
results are presented in this section.

\subsection{\bf Type Ia supernovae}
The cosmic acceleration was discovered 16 years ago by measuring
accurate distances to distant SNe
Ia~\cite{Riess98,Schmidt98,Perlmutter99}. The reason for the
acceleration remains uncertain and a large experimental effort in
observational cosmology has been driven to reveal the mechanism of
this ostensibly counterintuitive phenomenon. By precisely mapping
the distance-redshift relation up to redshift $z\approx1$, SNe Ia
remain, at this stage, the most promising probe of the late-time
history of the Universe. Because of the variability of the large
spectra features, distance estimation for SNe Ia is based on the
empirical observation that these events form a homogeneous class
whose remaining variability is reasonably well captured by two
parameters~\cite{Tripp98}. One of them characterizes the stretching
of the light curve ($X_1$ in what follows), and the other describes
the color at maximum brightness ($\mathcal{C}$ in what follows).

With the assumption that SNe Ia at all redshifts with the identical
color, shape and galactic environment have, on average, the same
intrinsic luminosity, the distance estimator (distance modulus:
$\mu=5\log\big[\frac{d_\mathrm{L}}{\mathrm{Mpc}}\big]+25$) used in
most cosmological analysis is quantified by a linear model,
\begin{equation}\label{eq10}
\mu_\mathrm{B}(\alpha, \beta; M)=m_\mathrm{B}^*-M+\alpha\times
X_1-\beta\times\mathcal{C},
\end{equation}
where $m_\mathrm{B}^*$ is the observed peak magnitude in the
rest-frame $B$ band, and $\alpha$ and $\beta$ are nuisance
parameters which characterize the stretch-luminosity and
color-luminosity relationships, corresponding to the well-known
broader-brighter and bluer-brighter relationships, respectively. The
value of $M$ is another nuisance parameter representing the absolute
magnitude of a fiducial SNe Ia. In general, $\alpha$ and $\beta$ are
left as free parameters (on the same weight as cosmological
parameters) that are determined in the global fit in the context of
standard dark energy scenario to construct the Hubble diagram for
SNe Ia. It should be noted that cosmological implications derived
from this Hubble diagram for other nonstandard models, which are
different from the standard $\Lambda$CDM (or $w$CDM) scenario used
to carry out the global fit, are model biased.

In order to achieve model-unbiased constraints on the clumpiness of
the Universe, we should fit the light-curve fitting parameters
($\alpha$ and $\beta$) and the model parameters ($\Omega_m$ and
$\nu$) simultaneously to construct a Hubble diagram of SNe Ia in an
inhomogeneity-allowed scenario by confronting the distances
estimated from SNe Ia observations via Eq.~(\ref{eq10}) with the
ones predicted from the ZKDR luminosity distance model,
\begin{equation}\label{eq11}
\mu_{\mathrm{mod}}(z;\mathbf{\theta_1},\mu_0)=5\log_{10}[D_\mathrm{L}(z;\mathbf{\theta_1})]+\mu_0.
\end{equation}
Here $D_\mathrm{L}$ is the Hubble-constant free luminosity distance,
$\mathbf{\theta_1}$ represents the model parameter vector
$(\Omega_m,\nu)$ and $\mu_0=5\log_{10}[c/H_0]+25$. For the latest
JLA SN Ia, the standard $\chi^2$ function is given by
\begin{equation}\label{eq12}
\chi^2(\mu_0,M;\mathbf{\theta_1},\mathbf{\theta_2})=\sum_{i=1}^{740}\frac{[\mu_{\mathrm{mod}}(z_i;
\mathbf{\theta_1},\mu_0)-\mu_{\mathrm{B},i}(\mathbf{\theta_2};M)]^2}{\sigma_{\mu,i}^2},
\end{equation}
where $\mathbf{\theta_2}$ denotes the vector of light-curve fitting
parameters $(\alpha,\beta)$ and $\sigma_{\mu,i}$ is the error on the
distance modulus for the $i$th SNe Ia. It should be noted that we
take only the statistical uncertainties into account and they are
also dependent on the light-curve fitting parameters. In order to
marginalize over the nuisance parameters, $H_0$ and $M$, we expand
the $\chi^2$ function with respect to $\widetilde{\mu_0}=\mu_0+M$
as~\cite{Pietro03,Nesseris05,Perivolaropoulos05}
\begin{equation}\label{eq13}
\chi^2(\mathbf{\theta_1},\mathbf{\theta_2};\widetilde{\mu_0})=A-2\widetilde{\mu_0}B+\widetilde{\mu_0}^2C,
\end{equation}
where
\begin{eqnarray}\label{eq14}
A(\mathbf{\theta_1},\mathbf{\theta_2})&=&\sum_{i=1}^{740}\frac{[\mu_{\mathrm{mod}}(z_i;
\mathbf{\theta_1},\mu_0=0)-\mu_{\mathrm{B},i}(\mathbf{\theta_2};M=0)]^2}{\sigma_{\mu,i}^2}\;,\\
B(\mathbf{\theta_1},\mathbf{\theta_2})&=&\sum_{i=1}^{740}\frac{[\mu_{\mathrm{mod}}(z_i;
\mathbf{\theta_1},\mu_0=0)-\mu_{\mathrm{B},i}(\mathbf{\theta_2};M=0)]}{\sigma_{\mu,i}^2}\;,\\
C(\mathbf{\theta_2})&=&\sum_{i=1}^{740}\frac{1}{\sigma_{\mu,i}^2}\;.
\end{eqnarray}
Equation~(\ref{eq13}) has a minimum at $\widetilde{\mu_0}=B/C$, and
it is
\begin{equation}\label{eq15}
\widetilde{\chi}^2(\mathbf{\theta_1},\mathbf{\theta_2})=A-\frac{B^2}{C}.
\end{equation}
Therefore, we can minimize
$\widetilde{\chi}^2(\mathbf{\theta_1},\mathbf{\theta_2})$ to get rid
of the dependence on nuisance parameters.

\begin{figure}[htbp]
\centering
\includegraphics[width=0.60\textwidth, height=0.45\textwidth]{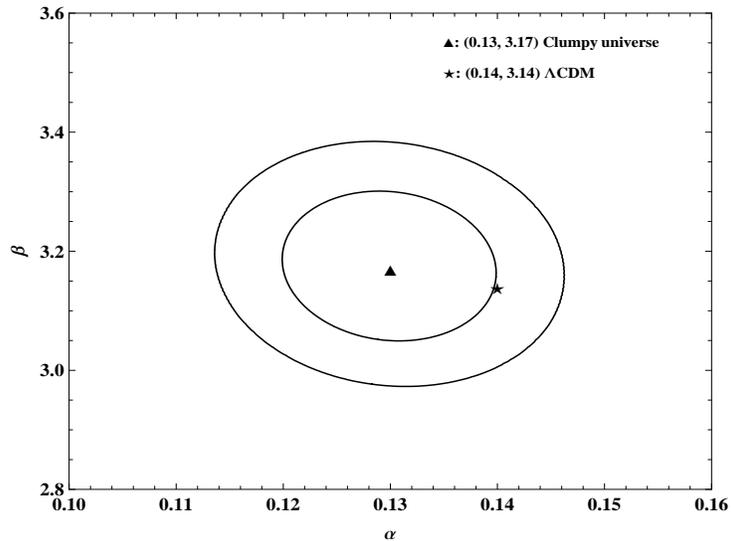}
\caption{\label{Fig1} Constraints on the light-curve fitting
parameters, $\alpha$ and $\beta$, from the global fit in the context
of a clumpy Universe. The triangle and star represent the best fits
when the ZKDR approximation and the standard $\Lambda$CDM framework
are considered, respectively.}
\end{figure}

\begin{figure}[htbp]
\centering
\includegraphics[width=0.60\textwidth, height=0.45\textwidth]{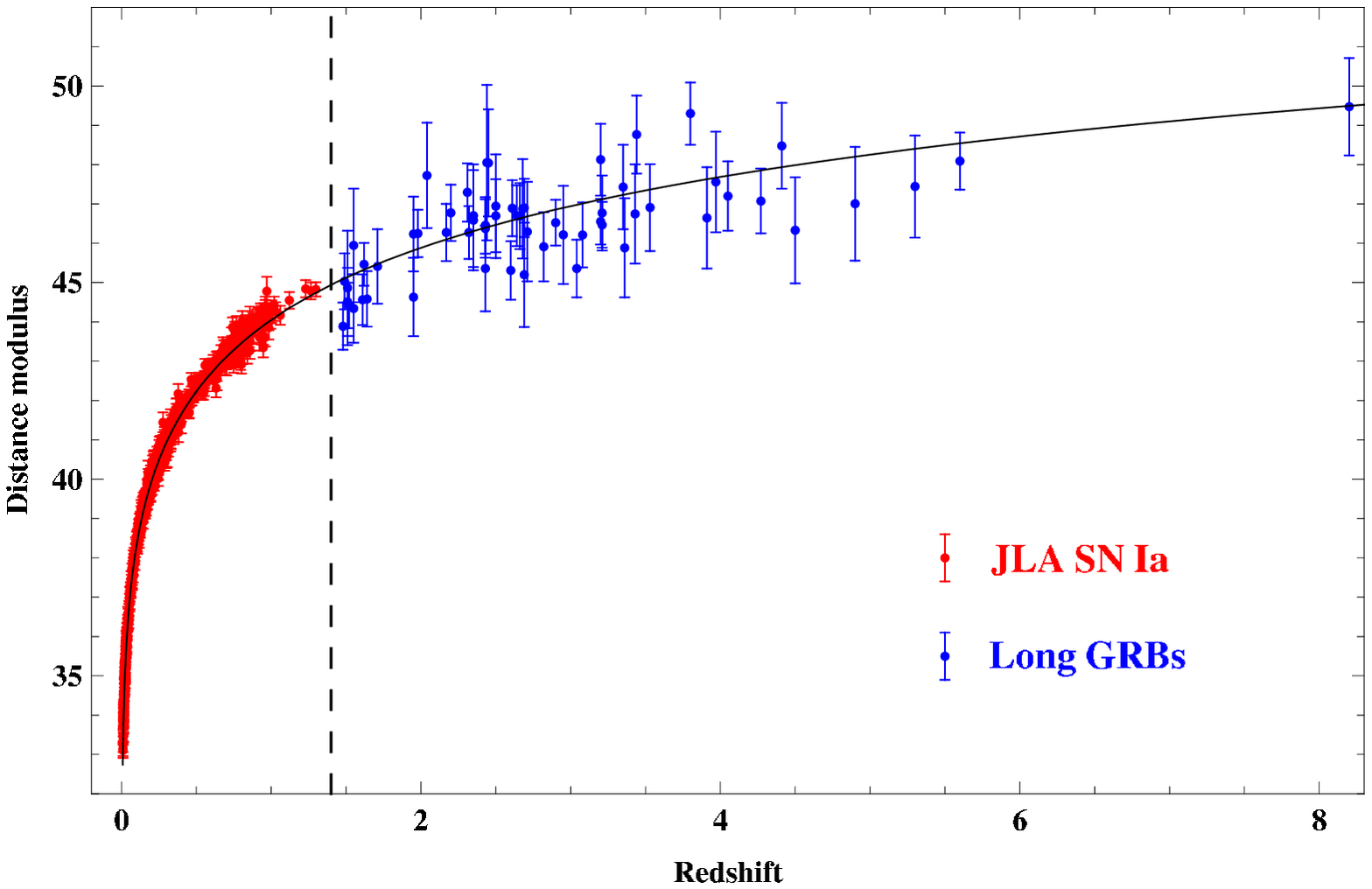}
\caption{\label{Fig2} Hubble diagram of the standard candles
constructed from the global fit in the context of a clumpy Universe.
The distance modulus redshift relation of the best-fit ZKDR
approximation for a fixed
$H_0=70~\mathrm{km}~\mathrm{s}^{-1}~\mathrm{Mpc}^{-1}$ is shown as
the solid line.}
\end{figure}

\begin{figure}[htbp]
\centering
\includegraphics[width=0.60\textwidth, height=0.45\textwidth]{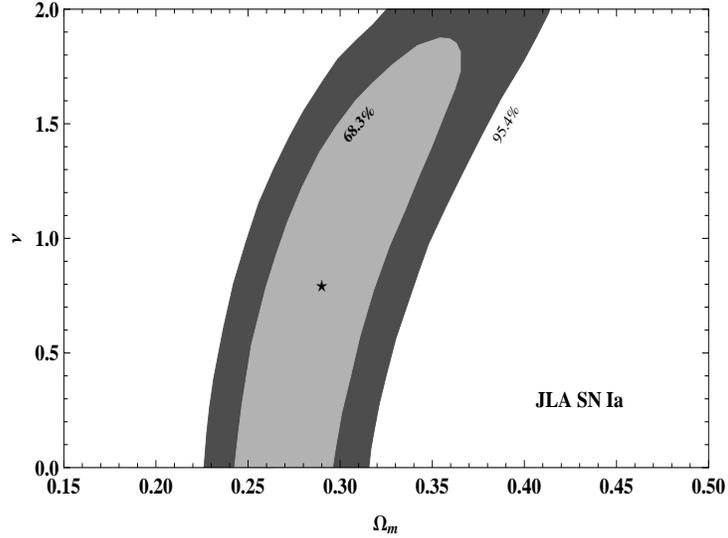}
\caption{\label{Fig3} Confidence regions in the ($\Omega_m$, $\nu$)
plane for the model with a ZKDR luminosity distance constrained from
the JLA SN Ia.}
\end{figure}

The constraint on the light-curve fitting parameters vector is
presented in Fig.~\ref{Fig1}. The best fit value is $(\alpha,~
\beta)=(0.13,~3.17)$, which is marginally compatible with the result
estimated in the flat $\Lambda$CDM at a 1$\sigma$ confidence level.
By applying a minimization of $\widetilde{\chi}^2$, we can get an
estimation for $\widetilde{\mu_0}$ which is a combination of $H_0$
and $M$. Here, we break the degeneracy by fixing
$H_0=70~\mathrm{km~s}^{-1}~\mathrm{Mpc}^{-1}$ and obtain $M=-19.08$.
With the constraint on ($\alpha,~\beta$) and estimation of $M$, an
indicative Hubble diagram in the framework of the ZKDR luminosity
distance model is constructed and shown in Fig.~\ref{Fig2}.
Moreover, results for confidence regions constrained in the
($\Omega_m,~\nu$) plane are presented in Fig.~\ref{Fig3} and
Tab.~\ref{Tab2}. We suggest that the clumpiness parameter $\eta$ is
poorly constrained, being bounded on the interval
$0.16\leq\eta\leq1.00$ within a 1$\sigma$ confidence level. However,
a tighter constraint is obtained for the matter density parameter
$\Omega_m$, being restricted on the interval
$0.25\leq\Omega_m\leq0.37$(1$\sigma$). These are very similar to
what was obtained in previous analyses~\cite{Santos08a,Busti12}, but
quite different from the results included in Ref.~\cite{Breto13}.
That is, our unbiased tests slightly indicate an inhomogeneity and
the standard FLRW cosmology is consistent with SNe Ia observations
within a 1$\sigma$ confidence level.

\subsection{\bf Long gamma-ray bursts}
Gamma-ray bursts (GRBs), which are the most intensive explosions
observed in the Universe and thus are visible across much larger
distances than SNe Ia, are deemed as a potential probe to explore
the Universe at higher redshift, a redshift of at least $6$ and up
to even $z=10$~\cite{Lamb2000, Bromm2002, Lin2004, Krimm2006}.
Specifically, relations between the luminosity/energy and the
measurable properties of the prompt gamma-ray emission imply that
GRBs may be appropriate candidates for cosmological standard
candles. In the past few years, several empirical luminosity
relations have been statistically inferred from observations. For
instance, several two-variable relations: the relation between
spectral lag and luminosity
($\tau_{\mathrm{lag}}-L$)~\cite{Norris2000}, the relation between
variability and luminosity ($V-L$)~\cite{Fenimore2000,
Reichart2001}, the relation between peak spectral energy and
luminosity ($E_\mathrm{peak}-L$)~\cite{Schaefer2003, Yonetoku2004},
the relation between peak spectral energy and collimation-corrected
energy ($E_\mathrm{peak}-E_\gamma$)~\cite{Ghirlanda2004}, the
relation between the minimum raising time in the GRB light curve and
luminosity ($\tau_{\mathrm{RT}}-L$)~\cite{Schaefer2007}, and the
relation between peak spectral energy and isotropic energy
($E_{\mathrm{peak}}-E_\mathrm{\gamma,iso}$)~\cite{Amati2002}--have
been successfully deduced from observations. Meanwhile, a few
multivariable relations have also been obtained, such as the
connection between $E_{\mathrm{iso}}$, $E_\mathrm{peak}$, and the
break time of the optical afterglow light curves
($t_\mathrm{b}$)~\cite{Liang2005}, the correlation between the
luminosity, $E_\mathrm{peak}$, and the rest-frame ``high-signal"
time scale ($T_{0.45}$)~\cite{Firmani2006}. Moreover, these
luminosity relations have been proposed to calibrate GRBs as
distance indicators (see, e.g.,
Refs.~\cite{Ghirlanda2006a,Schaefer2007} for reviews).

In particular, in Refs.~\cite{Busti12,Breto13}, distances of GRBs
used to constrain the clumpiness of the Universe are obtained by
calibrating their luminosity relations with low-redshift SNe
Ia~\cite{Liang08,Hao10,Tsutsui12}. However, it is necessary to make
clear that distances of SNe Ia quoted to calibrate luminosity
relations are estimated from a global fit in the frame of a standard
dark energy model. In other words, the distances of GRBs given in
Refs.~ \cite{Liang08,Hao10,Tsutsui12} are still somewhat dependent
on the standard dark energy model and thus subsequent tests for the
inhomogeneity of the Universe derived from them are model biased. In
this work, we construct the Hubble diagram of 116 long
GRBs~\cite{Fayin11} in the framework of an inhomogeneous Universe by
calibrating their luminosity/energy relations in the global fit
where the context of the ZKDR luminosity distance model is
considered. This Hubble diagram can then lead to an unbiased
examination of the clumpiness of the Universe. In
Ref.~\cite{Fayin11}, six luminosity correlations
($\tau_{\mathrm{lag}}-L$, $V-L$, $E_{\mathrm{peak}}-L$,
$E_{\mathrm{peak}}-E_{\gamma}$, $\tau_{\mathrm{RT}}-L$,
$E_{\mathrm{peak}}-E_{\gamma, \mathrm{iso}}$) have been derived from
the latest observations of 116 long GRBs. In their work, it was also
found that the intrinsic scatter of the $V-L$ correlation was too
large to infer an inherent correlation between these two quantities
using the currently observed GRB events. What is more, the
luminosity correlations $E_{\mathrm{peak}}-E_{\gamma}$ and
$E_{\mathrm{peak}}-E_{\gamma, \mathrm{iso}}$ mirror almost the same
physics, we should include one of them to avoid strong correlation
among the luminosity correlations. Therefore, we choose the
$E_{\mathrm{peak}}-E_{\gamma}$ correlation, which has a smaller
intrinsic scatter, and then use the rest four correlations for the
following analysis. The same as previous works that derived
cosmological implications from GRBs, we use only the subsample at
$z>1.4$ for the complimentary redshift range to the SN Ia.

The remaining four luminosity correlations involved in this paper
are
\begin{align}
  \label{eq:GRB-lag-L}
  \log \frac{L}{1 \; \mathrm{erg} \; \mathrm{s}^{-1}}
  &= a_1+b_1 \log
  \left[
    \frac{\tau_{\mathrm{lag}}(1+z)^{-1}}{0.1\;\mathrm{s}}
  \right]
  ,
  \end{align}
  \begin{align}
  \label{eq:GRB-E_peak-L}
  \log \frac{L}{1 \; \mathrm{erg} \; \mathrm{s}^{-1}}
  &= a_2+b_2 \log
  \left[
    \frac{E_{\mathrm{peak}}(1+z)}{300\;\mathrm{keV}}
  \right]
  ,
   \end{align}
  \begin{align}
  \label{eq:GRB-E_peak-E_gamma}
  \log \frac{E_{\gamma}}{1\;\mathrm{erg}}
  &= a_3+b_3 \log
  \left[
    \frac{E_{\mathrm{peak}}(1+z)}{300\;\mathrm{keV}}
  \right]
  ,
  \end{align}
  \begin{align}
  \label{eq:GRB-tau_RT-L}
  \log \frac{L}{1 \; \mathrm{erg} \; \mathrm{s}^{-1}}
  &= a_4+b_4 \log
  \left[
    \frac{\tau_{\mathrm{RT}}(1+z)^{-1}}{0.1\;\mathrm{s}}
  \right]
  ,
  \end{align}

where $a$ and $b$ are the intercept and the slope of the relation,
respectively. In these correlations, the isotropic peak luminosity
$L$ is given by
\begin{equation}\label{eq:GRB-L-P_bolo}
L=4 \pi d_L^2 P_{\mathrm{bolo}},
\end{equation}
where $P_{\mathrm{bolo}}$ is the bolometric flux of gamma rays in
the burst. The isotropic energy released in a burst is
\begin{equation}\label{eq:GRB-E_iso-S_bolo}
E_{\gamma,\mathrm{iso}}=4 \pi d_L^2 S_{\mathrm{bolo}} (1+z)^{-1},
\end{equation}
where $S_{\mathrm{bolo}}$ is the bolometric fluence of gamma rays in
the burst at redshift $z$. The total collimation-corrected energy
can be calculated by
\begin{equation}\label{eq:GRB-E_gamma-S_bolo}
E_{\gamma}= E_{\gamma,\mathrm{iso}} (1-\cos\theta_{\mathrm{jet}}),
\end{equation}
where $\theta_{\mathrm{jet}}$ is the opening angle of the jet.

\begin{figure}[htbp]
\centering
\includegraphics[width=0.60\textwidth, height=0.45\textwidth]{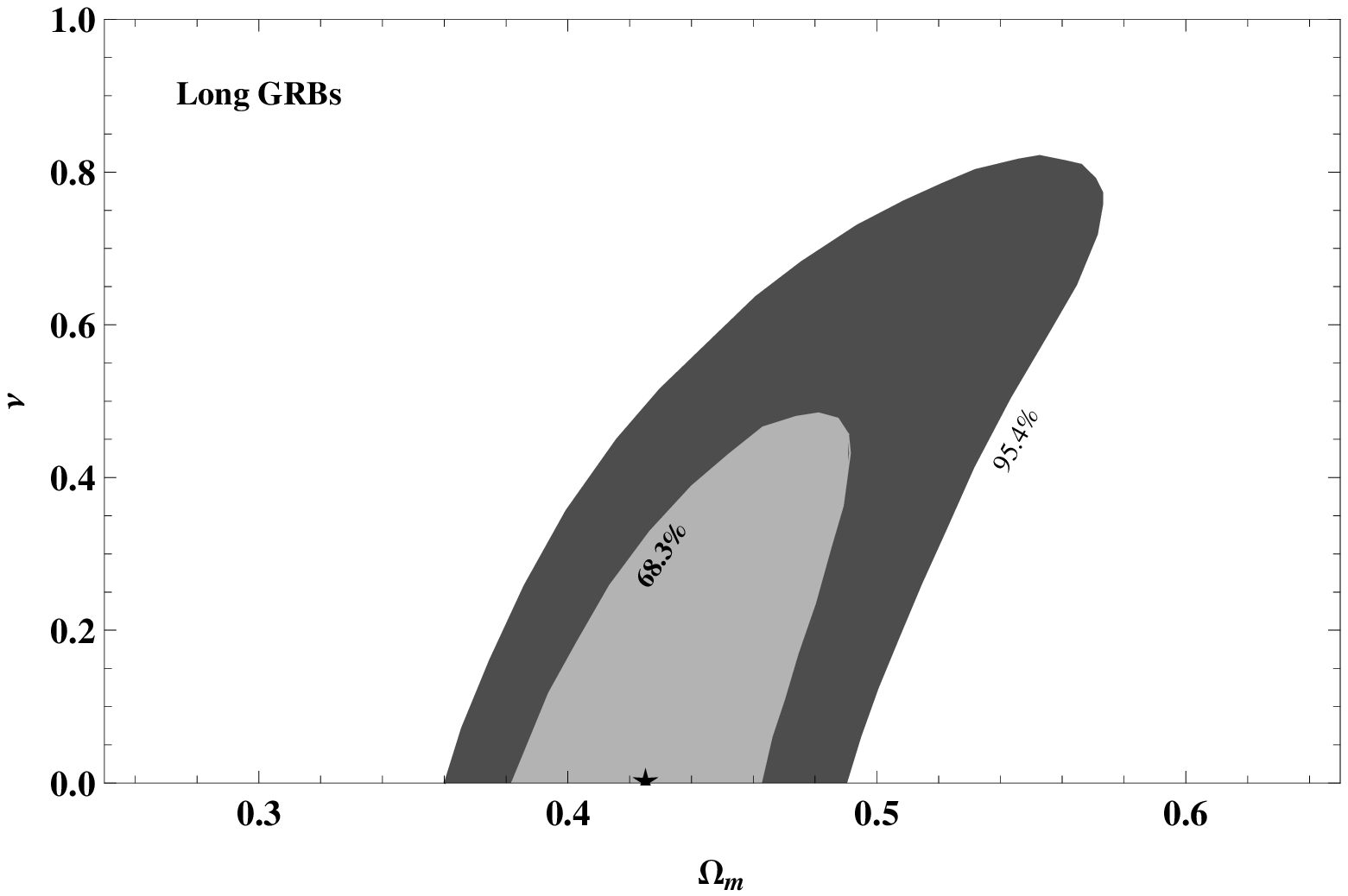}
\caption{\label{Fig4} Confidence regions in the ($\Omega_m$, $\nu$)
plane for the model with a ZKDR luminosity distance constrained from
the long GRBs.}
\end{figure}

In order to completely avoid any circularity and obtain
model-unbiased constraints on the clumpiness of the Universe from
GRBs~\cite{Schaefer2003,Schaefer2007}, we separately calibrate each
luminosity relation, Eqs.~\ref{eq:GRB-lag-L}-\ref{eq:GRB-tau_RT-L},
by carrying out a similar simultaneous global fitting route
presented in the above subsection. Results are shown in
Tab.~\ref{Tab1}. Here, $\sigma_{\mathrm{int}}$ is the systematic
error and it can be estimated by finding the value such that an
$\chi^2$ fit to each relation calibration curve produces a value of
reduced $\chi^2$ of unity~\cite{Schaefer2007}. This quantity
accounts the extra scatter of the luminosity relations. In this
global fitting route, we marginalize the nuisance parameter Hubble
constant by fixing $H_0=70~\mathrm{km~s}^{-1}~\mathrm{Mpc}^{-1}$.
Following the method about uncertainty calculation and distance
estimation from calibrated luminosity
relations~\cite{Schaefer2007,Liang08}, as shown in Fig.~\ref{Fig2},
we construct a Hubble diagram of GRBs in the context of the ZKDR
luminosity distance scenario. In addition, results concerning the
constraints on model parameters are presented in Fig.~\ref{Fig4} and
Tab.~\ref{Tab2}. It is suggested that a Universe composed only by
homogeneously distributed matter is strongly favored by GRB
observations. This is greatly different from what was obtained in
previous works~\cite{Busti12,Breto13}.

\begin{figure}[htbp]
\centering
\includegraphics[width=0.60\textwidth, height=0.45\textwidth]{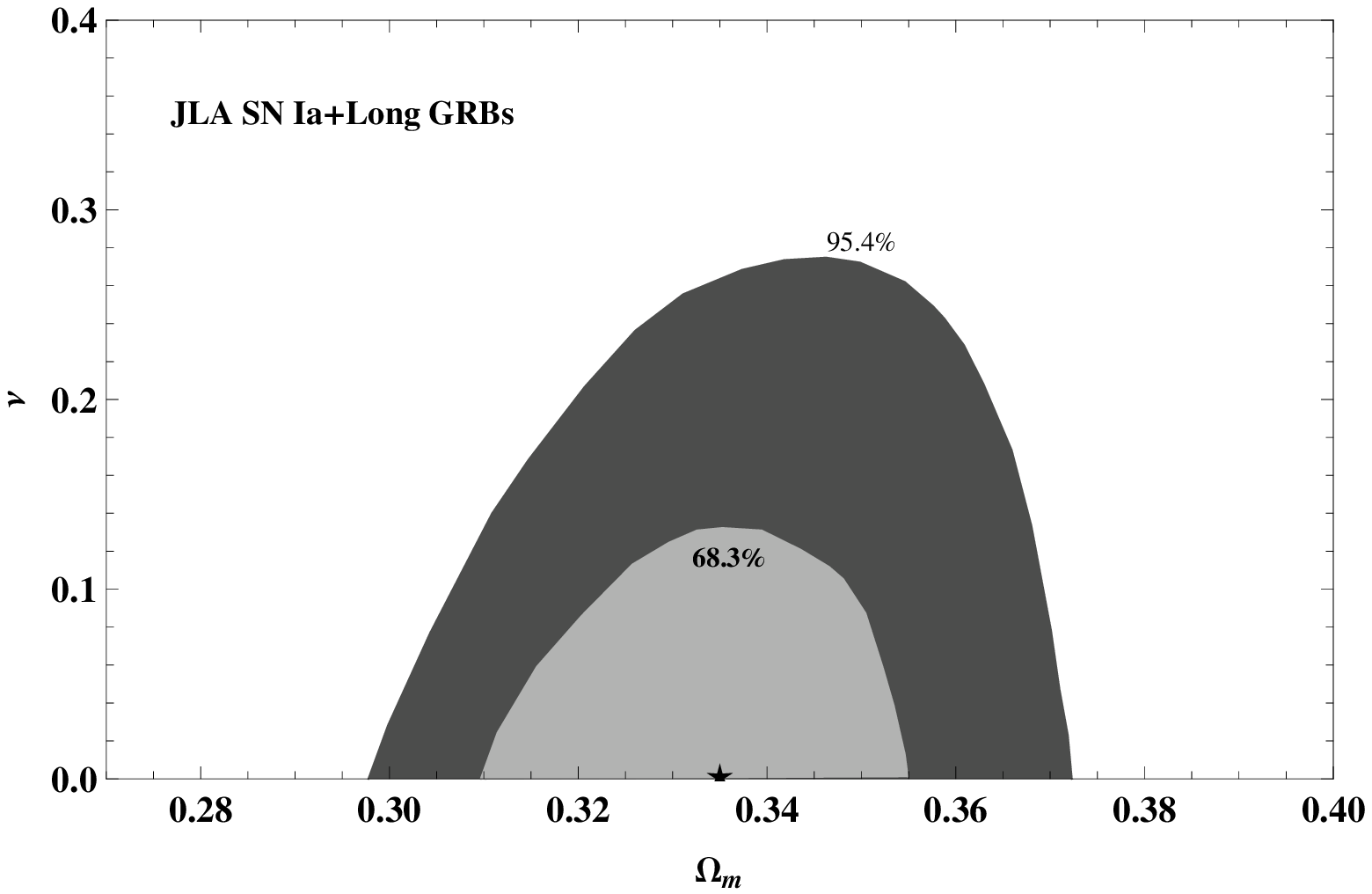}
\caption{\label{Fig5} Confidence regions in the ($\Omega_m$, $\nu$)
plane for the model with a ZKDR luminosity distance constrained from
the combination of JLA SN Ia and long GRBs.}
\end{figure}

Finally, we perform a joint analysis from the combination of JLA SN
Ia and long GRBs. Results are displayed in Fig.~\ref{Fig5} and
Tab.~\ref{Tab2}. Within a 1$\sigma$ confidence level, the matter
density parameter is restricted in the interval
$0.32\leq\Omega_m\leq0.36$ and the smoothness parameter is bounded
in the interval $0.98\leq\eta\leq1.00$. It is shown that the
constraint on the matter density parameter is mainly dependent on
SNe Ia observations while the estimation of the smoothness parameter
is basically determined by the long GRBs. The fact that high
redshift GRBs prefer a homogeneous Universe with a great
significance of probability can be understood as follows: they
explore much larger scales of the Universe and should contribute to
diminishing the corresponding space parameter. That is, since the
Universe is more homogeneous on larger scales (a higher redshift),
higher value of the smoothness parameter $\eta$ is favored. In
addition, it should be noted that, although large redshift GRBs are
very important for the tests of the clumpiness parameter, there are
only four GRBs at redshift larger than $5$.

\begin{table}[!h]
\begin{tabular}{|c|c|c|c|c|}
\hline
~Luminosity relation ~&~~$a(1\sigma)$~~&~~$b(1\sigma)$ ~~&~~$\sigma_{\mathrm{int}}$~~&~~$N(\mathrm{z}_{\mathrm{GRB}}>1.4)$~~\\
\hline
$\tau_{\mathrm{lag}}-L$  &~~ $52.60\pm0.04$~~&~~$-0.76\pm0.06$~~&~~0.12~~&~~26~~\\
\hline
$E_{\mathrm{peak}}-L$  &~~ $52.10\pm0.04$~~&~~$1.38\pm0.12$~~&~~0.16~~&~~62~~\\
\hline
$E_{\mathrm{peak}}-E_\gamma$  &~~ $50.36\pm0.07$~~&~~$1.56\pm0.20$~~&~~0.01~~&~~12~~\\
\hline
$\tau_{\mathrm{RT}}-L$  &~~ $52.95\pm0.05$~~&~~$-1.03\pm0.13$~~&~~0.16~~&~~36~~\\
\hline
\end{tabular}
\tabcolsep 0pt \caption{\label{Tab1} Summary of the constraints on
luminosity relations of GRBs from the global fit in the context of a
clumpy Universe.} \vspace*{5pt}
\end{table}

\begin{table}[!h]
\begin{tabular}{|c|c|c|c|}
\hline
~Sample ~&~~$\Omega_m(1\sigma)$~~&~~$\nu(1\sigma)$ ~~&~~$\eta(1\sigma)$~~\\
\hline
~~JLA SN Ia~~&~~ $0.25\leq\Omega_m\leq0.37$~~&~~$0.00\leq\nu\leq1.80$~~&~~$0.16\leq\eta\leq1.00$~~\\
\hline
~~Long GRBs~~&~~ $0.38\leq\Omega_m\leq0.49$~~&~~$0.00\leq\nu\leq0.48$~~&~~$0.88\leq\eta\leq1.00$~~\\
\hline
~~Joint analysis~~&~~ $0.32\leq\Omega_m\leq0.36$~~&~~$0.00\leq\nu\leq0.12$~~&~~$0.98\leq\eta\leq1.00$~~\\
\hline
\end{tabular}
\tabcolsep 0pt \caption{\label{Tab2} Summary of the unbiased
constraints on model parameters in the ZKDR luminosity distance from
observations of standard candles. } \vspace*{5pt}
\end{table}

\section{CONCLUSIONS AND DISCUSSIONS}
In the era of precision cosmology, where one aims at determining the
cosmological parameters at the percent level, distance estimations
for standard candles and rulers with increasing accuracy are
expected to provide powerful constraints on dark energy or other
fundamental dynamical parameters. However, it is necessary to be
aware of the physical hypothesis underlying these probes when we
proceed with such a program. As far as we know, the Universe is
effectively inhomogeneous at least in the small-scale domain.
Furthermore, notice that even the large-scale homogeneity also has
been challenged~\cite{Uzan08}. In this topic, the method based on
the ZKDR luminosity distance is a simple alternative and is usually
applied to quantitatively assessing the influences of the clumpiness
on the light propagation. In the past few years, there has been a
rich literature concerning the constraints on the smoothness
parameter from observations of standard
candles~\cite{Kantowski01,Santos08b,Busti12,Breto13,Lima14}.
However, we should keep in mind that distances of SNe Ia applied to
test the inhomogeneity were estimated from the global fit in the
context of a standard homogeneous dark energy model, i.e., the flat
$\Lambda$CDM or $w$CDM model. Therefore, in these previous analyses,
constraints on the smoothness parameter from the distance modulus of
SNe Ia were somewhat model biased. Meanwhile, results obtained from
GRBs suffered the same problem since the distances of them were
determined by calibrating luminosity relations with low-redshift SNe
Ia.

In this paper, we first construct Hubble diagrams for SNe Ia and
GRBs by calibrating the light-curve fitting parameters and
luminosity relations, respectively, in the global fit where the
context of the ZKDR luminosity distance model is taken into account.
And then, these Hubble diagrams can lead to unbiased tests for the
inhomogeneity of the Universe. For the JLA SN Ia, as shown in
Fig.~\ref{Fig3}, constraint on the smoothness parameter is not
stringent and slightly implies a locally inhomogeneous background,
while the matter density parameter is well constrained, being
bounded in the interval $0.25\leq\Omega_m\leq0.37$(1$\sigma$). For
the long GRBs, as shown in Fig.~\ref{Fig4}, the Universe with matter
uniformly distributed is favored with a high confidence level. This
is completely different from what was obtained in
Ref.~\cite{Breto13}. Finally, we perform a joint analysis which
provides good constraints on both model parameters. At a 1$\sigma$
confidence level, the intervals are $0.32\leq\Omega_m\leq0.36$ and
$0.98\leq\eta\leq1.00$. It is suggested that the constraint on the
matter density parameter is mainly based on the observations of
low-redshift SNe Ia, while the test for the clumpiness parameter is
primarily determined from the observations of high-redshift GRBs.
Just as expected, the investigation on the inhomogeneity was very
sensitive to the scales explored by the observations, i.e., the
Universe should be more homogeneous on larger scales. These also may
be an indication that the ZKDR approximation remains to be a precise
description for the luminosity distance-redshift relation in a
locally inhomogeneous Universe with the cosmological constant.

Frankly, it should be pointed out that constraints on the model
parameters from low-redshift SNe Ia and high-redshift GRBs are
somewhat inconsistent. This inconsistency may imply that the
assumption with the smoothness parameter $\eta$ being a constant is
not accurate enough to fit the practical observations. That is, the
smoothness parameter $\eta$ might evolve with cosmic time (or
redshift). Moreover, the intrinsic scatters in GRB observations may
also lead to this tension. Therefore, in the near future, a more
precise and larger sample of high-redshift GRB data (even some other
distance measurements with new methods, e.g., extremely luminous
active galactic nuclei readily observed over a range of distances
from $\sim10~\mathrm{Mpc}$ to $z>7$~\cite{Watson2011, Wang2013,
Yoshii2014}) and a plausible extension of the ZKDR approach are
expected to perform more accurate tests for the inhomogeneity and
contribution of matter in the Universe.

\section*{Acknowledgments}
We are grateful to the anonymous referee for his or her helpful
comments. This work was supported by the Ministry of Science and
Technology National Basic Science Program (Project 973) under Grants
No. 2012CB821804 and No. 2014CB845806, the Strategic Priority
Research Program ``The Emergence of Cosmological Structure" of the
Chinese Academy of Sciences (No. XDB09000000), the National Natural
Science Foundation of China under Grants No. 11373014 and No.
11073005, the China Postdoc Grant No. 2014T70043, and the
Fundamental Research Funds from the Central Universities and
Scientific Research Foundation of Beijing Normal University.

\end{document}